\documentclass[conference,table,usenames,dvipsnames,svgnames]{IEEEtran}
\pdfoutput=1  

\usepackage[T1]{fontenc}

\usepackage[style=ieee,backend=biber,bibencoding=utf8,texencoding=ascii]{biblatex}
\usepackage{amsmath}
\usepackage{amssymb}
\usepackage{braket}
\usepackage{bm}
\usepackage{cases}  
\usepackage[binary-units]{siunitx}

\usepackage[basic]{complexity}
\newclass{\MAXSAT}{MaxSAT}
\newclass{\MAXCUT}{MaxCUT}

\usepackage{graphicx}
\usepackage[font=footnotesize]{caption}
\usepackage{subcaption}
\usepackage{hyperref}
\usepackage[capitalise]{cleveref} 
\usepackage{csquotes}
\usepackage{booktabs}

\usepackage{float}
\usepackage{xspace}

\DeclareCaptionLabelFormat{nocaption}{}

\title{Portfolio rebalancing experiments using the Quantum Alternating Operator Ansatz}

\author{
	\IEEEauthorblockN{
		Mark Hodson\IEEEauthorrefmark{1},
		Brendan Ruck\IEEEauthorrefmark{1},
		Hugh (Hui Chuan) Ong\IEEEauthorrefmark{2},
		David Garvin\IEEEauthorrefmark{1},
		Stefan Dulman\IEEEauthorrefmark{2}
	}
	\IEEEauthorblockA{
		\IEEEauthorrefmark{1}Rigetti Computing
	}
	\IEEEauthorblockA{
		\IEEEauthorrefmark{2}Commonwealth Bank of Australia
	}
}

\begin{document}


\IEEEtitleabstractindextext{%
	\begin{abstract}
		This paper investigates the experimental performance of a discrete portfolio optimization problem relevant to the financial services industry on the gate-model of quantum computing. We implement and evaluate a portfolio rebalancing use case on an idealized simulator of a gate-model quantum computer. The characteristics of this exemplar application include trading in discrete lots, non-linear trading costs, and the investment constraint. We design a novel problem encoding and hard constraint mixers for the Quantum Alternating Operator Ansatz, and compare to its predecessor the Quantum Approximate Optimization Algorithm. Experimental analysis demonstrates the potential tractability of this application on Noisy Intermediate-Scale Quantum (NISQ) hardware, identifying portfolios within 5\% of the optimal adjusted returns and with the optimal risk for a small eight-stock portfolio.
	\end{abstract}
}

\maketitle

\IEEEdisplaynontitleabstractindextext
\IEEEpeerreviewmaketitle



\section{Introduction}
\label{sec:introduction}

Gate-model Noisy, Intermediate-Scale Quantum (NISQ) \cite{Preskill2018} computers are becoming increasingly available in the cloud, and of sufficient scale and fidelity to run interesting quantum algorithms. Quantum algorithms such as the Quantum Approximate Optimization Algorithm of \cite{Farhi2014} and the Quantum Alternating Operator Ansatz of \cite{Hadfield2019}, collectively QAOA, are able to execute on NISQ hardware and provide the potential of advantage at larger scales. Mapping of industry applications onto quantum algorithms has begun, including in financial services \cite{Orus2018}, and questions are being raised as to the relevance, tractability and performance of quantum approaches to financial services problems in the quantum computing ecosystem.

In this paper we have brought together financial services and quantum software technologists to select, implement, and test a portfolio rebalancing use case using QAOA. We designed and built software based on the knowledge discussed in this paper which demonstrates this application, and executed the software on an idealized simulator of a gate-model quantum computer. We describe this financial application and its relevance, its formulation in both QAOA variants, experimental results, lessons learned, and avenues for further development.


\section{Preliminaries}
\label{sec:preliminaries}

We summarize quantum algorithms and identities upon which this research is based.

\subsection{Binary to spin system identity}

Conversion from a binary system based on $x \in \lbrace 0, 1 \rbrace$ to a spin system based on $s \in \lbrace -1, +1 \rbrace$ is afforded by substitution using the identity

\begin{equation}
	\label{eq:binary_to_spin_system_identity}
	s = 2x-1
\end{equation}

\subsection{Penalty functions for soft constraints}

A real-valued ($\mathbf{x} \in \mathbb{R}^N$, $c \in \mathbb{R}$) equality constraint of the form

\begin{equation*}
	f(\mathbf{x}) = c
\end{equation*}

can be converted to a form that can be solved by unconstrained optimization using a \emph{penalty function} as

\begin{equation}
	\label{eq:penalty_functions_for_soft_constraints}
	P(\mathbf{x}) = A \left( f(\mathbf{x}) - c \right)^2
\end{equation}

where \begin{itemize}
	\item $P(\mathbf{x}) = 0$ when the constraint is met
	\item $P(\mathbf{x}) > 0$ when the constraint is violated
	\item $A \in \mathbb{R} \mid A > 0$ is a penalty scaling coefficient
\end{itemize}

\subsection{Quantum approximate optimization}
\label{sec:preliminaries_quantum_approximate_optimization}

The Quantum Approximate Optimization Algorithm \cite{Farhi2014} has been extended \cite{Wang2017} to minimize a polynomial cost function with real-valued coefficients and discrete solution variables. Discrete solution variables can be defined as a binary system $\mathbf{x} \in \lbrace 0, 1 \rbrace^N$ or spin system $\mathbf{s} \in \lbrace -1, +1 \rbrace^N$ of $N$ variables.

In its canonical form, QAOA's polynomial cost function is a sum-of-products expression with each product term interacting between $0$ and $N$ of the solution variables. If we consider each possible unique interaction then, using the binomial theorem, the maximum number of terms is $2^N$. We observe that the total number of non-zero polynomial coefficients in a tractable problem formulation must scale favorably with respect to the problem size, as each coefficient must be calculated during pre-processing and input as a parameter to the QAOA circuit.

Many interesting optimization problems have at most quadratic terms in the polynomial cost function \cite{Lucas2014}. We restrict ourselves to quadratic problems formulated as a spin system. This results in the Ising model optimization cost function familiar to quantum annealing, as

\begin{equation}
	\label{eq:ising_model_cost_function}
	C(\mathbf{s}) = c + \sum_{i=1}^N h_i s_i + \sum_{i=1}^N \sum_{j=i+1}^N J_{ij} s_i s_j
\end{equation}

where \begin{itemize}
	\item $c \in \mathbb{R}$ is a constant term
	\item $h_i \in \mathbb{R}$ is a coefficient of the \emph{bias vector} $\mathbf{h}$
	\item $J_{ij} \in \mathbb{R}$ is a coefficient of the upper-triangular \emph{coupling matrix} $J$
\end{itemize}

The execution of QAOA is as a \emph{variational algorithm} where circuit parameters $\mathbf{\beta} \in [ 0, \pi ]^p$ and $\mathbf{\gamma} \in [ 0, 2\pi ]^p$ are varied to minimize the expectation value $\langle \psi_1 | C | \psi_1 \rangle$, and so measure \enquote{good} solutions with high probability where

\begin{equation}
	\label{eq:qaoa_initial_state}
	| \psi_0 \rangle = | + \rangle ^ {\otimes N}
\end{equation}

is the initial state of the system, and

\begin{equation}
	\label{eq:qaoa_final_state}
	| \psi_1 \rangle = \left( \prod_{\alpha = p}^1 U(B, \beta_{\alpha}) \, U(C, \gamma_{\alpha}) \right) | \psi_0 \rangle
\end{equation}

is the final state of the system.

In \cref{eq:qaoa_final_state}, unitary evolution occurs via two exponentiated operators: $U(B, \beta_{\alpha}) = e^{-i \beta_{\alpha} B}$ and $U(C, \gamma_{\alpha}) = e^{-i \gamma_{\alpha} C}$, with $i = \sqrt{-1}$ being the imaginary number. The gates applied in a quantum computer iterate as $\alpha = 1, \cdots, p$ due to the right-associativity of these operations. Parameter $p \in \mathbb{N}$ is the number of parameterized repetitions in the resulting quantum circuit, and relates linearly to its depth. The quantum circuit hyper-parameter space of $\beta$ and $\gamma$ also increases linearly with $p$, and is optimized classically.

For our case of the Ising model cost function in \cref{eq:ising_model_cost_function}, the \emph{cost operator} is defined in the Pauli-Z basis ($\sigma^z$) as

\begin{equation}
	\label{eq:qaoa_ising_cost_operator}
	C = \sum_{i=1}^N h_i \sigma_i^z + \sum_{i=1}^N \sum_{j=i+1}^N J_{ij} \sigma_i^z \sigma_j^z
\end{equation}

and is not unitary but is Hermitian, allowing it to be used as the expectation value observable.

For unconstrained optimization problems, the Quantum Approximate Optimization Algorithm defines a \emph{mixing operator} that explores all $2^N$ combinatorial solutions. It is defined in the Pauli-X basis ($\sigma^x$) \cite[Equation (3)]{Farhi2014} in a way physically similar to quantum annealing, as

\begin{equation}
	\label{eq:qaoa_farhi_mixing_operator}
	B = \sum_{i=1}^N \sigma_i^x
\end{equation}

For constrained optimization problems, the Quantum Alternating Operator Ansatz suggests $B$ be designed to constrain the feasible subspace of solutions. One example of this is a \emph{parity mixer}, which uses alternating application of Pauli-XY mixers to odd and even spin subsets \cite[Equations (7)-(9)]{Hadfield2019}, we summarize as

\begin{equation}
	\label{eq:qaoa_parity_mixing_operator}
	U(B, \beta_{\alpha}) = U(B_{\text{last}}, \beta_{\alpha}) \, U(B_{\text{even}}, \beta_{\alpha}) \, U(B_{\text{odd}}, \beta_{\alpha})
\end{equation}

with

\begin{equation*}
	B_{\text{odd}} = \sum_{a \text{ odd}}^{N-1} \sigma_{a}^x \sigma_{a+1}^x + \sigma_{a}^y \sigma_{a+1}^y
\end{equation*}

\begin{equation*}
	B_{\text{even}} = \sum_{a \text{ even}}^N \sigma_{a}^x \sigma_{a+1}^x + \sigma_{a}^y \sigma_{a+1}^y
\end{equation*}

\begin{equation*}
	B_{\text{last}} = \Big\{ \begin{array}{l}
		\sigma_N^x \sigma_1^x + \sigma_N^y \sigma_1^y, \; N \text{ odd} \\[0.5em]
		I, \; N \text{ even}
	\end{array}
\end{equation*}

and where $I$ is the identity transform and all arithmetic is modulo $N$.

Such a mixer can, for example, be used to realize \emph{one-hot encoding} of categorical variables and has the potential to improve application performance over the original QAOA.

\section{Portfolio Rebalancing Application}
\label{sec:portfolio_rebalancing_application}

Portfolio rebalancing is a periodic asset management process in which traders maintain an institutional portfolio's net value, adjusting asset mix based on institutional advice and hedging risk as market conditions change. Rebalancing can be achieved using a combination of financial instruments including long and short positions in assets, and their derivatives (put and call options). The frequency of institutional rebalancing depends on many factors including market volatility, risk profile, asset liquidity and trading costs. It may be performed daily, weekly or monthly. The computational load of rebalancing, including the associated risk calculations, means that it is typically performed overnight. This creates an opportunity for new methods to significantly impact institutional process. For example, a significant speed up afforded by functional quantum computers could enable financial institutions to be more agile in their decision making, responding in a more timely way to changes in market conditions.

\subsection{Previous work}
\label{sec:portfolio_rebalancing_application_previous_work}

Discrete portfolio optimization of a Markowitz model portfolio \cite{Markowitz1952} is proven to be \NP{}-complete when described using integer positions \cite{Mansini1999}. The applicability of quantum computing to discrete portfolio optimization has been demonstrated using quantum annealing \cite{Rosenberg2016} and specialized gate-model algorithms that require quantum access to the historical record of returns \cite{Rebentrost2018}. In a recent review, portfolio optimization has not yet been evaluated using QAOA \cite[Section III]{Orus2018}. The potential for QAOA to provide guarantees on performance for problems such as \MAXCUT{} has been demonstrated \cite{Zhou2018}. The work of \cite{Hadfield2019} realizes hard constraints within QAOA using techniques available on today's gate-model NISQ computers. A similar concept has been proposed for quantum annealing \cite{Hen2016} but is not yet realized in hardware, meaning annealing solutions must use soft constraint emulations based on penalty functions \cite{Lucas2014}.

\subsection{Contribution}
\label{sec:portfolio_rebalancing_application_contribution}

Our contribution is the experimental evaluation of the tractability and performance of discrete portfolio optimization under constraints for a multi-period portfolio rebalancing scenario, implemented using QAOA on a simulator of a gate-model quantum computer. We design and implement a soft-constraint formulation based on \cite{Farhi2014}, and compare it to a hard-constraint formulation based on \cite{Hadfield2019}. We choose an idealized simulator of a gate-model quantum computer \cite{Smith2017} as a first step in understanding algorithm and application performance, and with the intent of evaluating on NISQ computers in the future. We choose portfolio rebalancing as an application of relevance to financial services institutions, drawn from our industry experience.

\subsection{Relevance}
\label{sec:portfolio_rebalancing_relevance}

Discrete portfolio optimization under constraints accurately models several characteristics of institutional trading in a portfolio rebalancing scenario.

The first characteristic is that of trading in discrete lots, where the asset class can only be traded in positive integer quantities, and the dollar value of a single lot is significant with respect to the total portfolio value. Asset classes such as bonds can have this property. In this situation, the closest integer position to the continuous form solution can be a poor approximation, and discrete optimization has the potential to gain a few basis points in the trading outcome.

The second characteristic is in modeling uncertainty. While a full treatment of uncertainty in the Markowitz model portfolio covariance matrix requires semi-definite programming and results in robust portfolio optimization \cite{Paini2018}, an approximation is to discretize the portfolio position space for each asset based on its own market variance. This provides some level of robustness to non-stationary market processes that are a known limitation in the Markowitz model.

The third characteristic is in representing the investment constraint, where the total portfolio value must be approximately maintained throughout the rebalancing process. This is common to the canonical Markowitz model discrete portfolio optimization problem \cite{Chan2016}, modified in a rebalancing scenario to consider the relative position with respect to previous, rather than the absolute position in the optimization process.

The fourth characteristic is in modeling trading costs, which are only accrued in the case that trading occurs on an asset. This non-linear function includes a step-change in cost which can be naturally represented with a discrete logic function. In our rebalancing scenario, we will assume that trading costs are fixed for any trade size.

Finally, other institutional trading practices can apply constraints to trading, such as the maximum number of assets that may be held\footnote{Commonly referred to as the cardinality constraint.}, minimum allocation sizes, and constraints on trading asset classes in a multi-class portfolio. These will not be expanded further. In this paper we will assume that a per-asset discrete lot size can be selected based on the first two characteristics presented here, and we will conduct a discrete portfolio optimization under constraints that includes trading costs and the investment constraint.


\section{Portfolio Rebalancing Formulation}
\label{sec:portfolio_rebalancing_formulation}

We define the problem of portfolio rebalancing to be a discrete portfolio optimization problem, as

\begin{equation}
	\label{eq:portfolio_rebalancing}
	\mathbf{z} = \underset{\mathbf{z}}{\mathrm{argmin}} \; C_{\text{RR}}(\mathbf{z}) + C_{\text{TC}}(\mathbf{z})
\end{equation}

where \begin{itemize}
	\item $\mathbf{z} = \lbrace -1, 0, +1 \rbrace^N$ is the solution vector of discrete portfolio asset positions to be held, representing long ($+1$), short ($-1$) or no-hold ($0$) states
	\item $C_{\text{RR}}(\mathbf{z})$ is the normalized risk-return function
	\item $C_{\text{TC}}(\mathbf{z})$ is the normalized trading cost function
	\item $N$ is the number of available portfolio assets
\end{itemize}

and subject to the investment constraint

\begin{equation}
	\label{eq:investment_constraint}
	\sum_{i=1}^N z_i = D
\end{equation}

where \begin{itemize}
	\item $D$ is the net total of discrete lots to be invested\footnote{Assuming the dollar value of each discrete asset lot is equal. In practice, trading lot sizes must be chosen carefully based on the criteria of \cref{sec:portfolio_rebalancing_relevance} and other institutional factors.}
\end{itemize}

The risk-return function is based on the Markowitz model \cite{Markowitz1952}, as

\begin{equation}
	\label{eq:risk_return_function}
	C_{\text{RR}}(\mathbf{z}) = \lambda \sum_{i=1}^N \sum_{j=1}^N \sigma_{ij} z_i z_j - \left( 1 - \lambda \right) \sum_{i=1}^N \mu_i z_i
\end{equation}

where \begin{itemize}
	\item $\lambda \in \mathbb{R} \mid 0 \leq \lambda \leq 1$ is an asset manager control parameter to favor low risk ($\lambda = 1$) or high return ($\lambda = 0$)
	\item $\sigma$ is the normalized asset returns covariance matrix
	\item $\mu$ is the normalized average asset returns vector
\end{itemize}

The trading cost function is based on a fixed trading cost model, as

\begin{equation}
	\label{eq:trading_cost_function}
    C_{\text{TC}}(\mathbf{z}) = \sum_{i=1}^N \left( 1 - \delta \left( z_i - y_i \right) \right) T
\end{equation}

where \begin{itemize}
	\item $\mathbf{y} = \lbrace -1, 0, +1 \rbrace^N$ is the previous portfolio position
	\item $\delta(z)$ is the discrete Dirac delta function, where $\delta(z) = 1$ if $z = 0$, and $\delta(z) = 0$ otherwise, with $z \in \mathbb{Z}$
	\item $T$ is the normalized cost incurred if an asset is traded
\end{itemize}

\subsection{Risk-return function encoding}
\label{sec:portfolio_rebalancing_formulation_risk_return_function_encoding}

We choose a two-spin encoding for $z_i \in \mathbf{z}$ by first defining independent binary decision variables for holding long and short positions, as

\begin{equation}
	\label{eq:long_short_decision_variables}
	z_i = x_i^+ - x_i^-
\end{equation}

where \begin{itemize}
	\item $x_i^+ \in \lbrace 0, 1 \rbrace$ decides upon a long position
	\item $x_i^- \in \lbrace 0, 1 \rbrace$ decides upon a short position
\end{itemize}

and then converting binary to spin variables as in \cref{eq:binary_to_spin_system_identity}

\begin{equation}
	\label{eq:risk_return_substitution}
	z_i = \frac{s_i^+ - s_i^-}{2}, \; (s_i^-, s_i^+) \in \lbrace -1, +1 \rbrace
\end{equation}

The four possible spin states are illustrated in \cref{tab:spin_encoding_of_portfolio_position_solution_variable}. The degenerate encoding of $x_i^- = 1, x_i^+ = 1$ will be penalized by the encoding of the trading cost function in \cref{sec:portfolio_rebalancing_formulation_trading_cost_function_encoding}.

\begin{table}
	\caption{Spin encoding of portfolio asset position variable $z_i$}
	\centering
	\begin{tabular}{ |c c|c c|c|c| }
		\hline
		$x_i^-$ & $x_i^+$ & $s_i^-$ & $s_i^+$ & $z_i$ & Description \\
		\hline
		$0$ & $0$ & $-1$ & $-1$ & $0$ & Flat (no position) \\
		$0$ & $1$ & $-1$ & $+1$ & $+1$ & Long position held \\
		$1$ & $0$ & $+1$ & $-1$ & $-1$ & Short position held \\
		$1$ & $1$ & $+1$ & $+1$ & $0$ & Netted off long and short \\
		\hline
	\end{tabular}
	\label{tab:spin_encoding_of_portfolio_position_solution_variable}
\end{table}

Substitution of the spin variable formulation \cref{eq:risk_return_substitution} into \cref{eq:risk_return_function} yields the spin-system risk-return cost function

\begin{equation}
	\label{eq:c_rr_s}
	\begin{alignedat}{2}
		& C_{\text{RR}}(\mathbf{s}) = \, & \lambda \sum_{i=1}^N \sum_{j=1}^N \frac{\sigma_{ij}}{4} (s_i^+ s_j^+ - s_i^+ s_j^- - s_i^- s_j^+ + s_i^- s_j^-) \\
		&                                & - (1 - \lambda) \sum_{i=1}^N \frac{\mu_i}{2} (s_i^+ - s_i^-)
	\end{alignedat}
\end{equation}

\subsection{Trading cost function encoding}
\label{sec:portfolio_rebalancing_formulation_trading_cost_function_encoding}

We design a spin-system trading cost function to realize the non-linear function of \cref{eq:trading_cost_function} over the domain of possible previous positions, of which there are three, and encoded new positions, of which there are four. The function is conditional on the previous position, as

\begin{equation}
	\label{eq:c_tc_s_conditional}
	C_{\text{TC}}(\mathbf{s}) = \Bigg\{ \begin{array}{l}
		\frac{1}{4} \, T \, (3 + s_i^+ - s_i^- + s_i^+ s_i^-) \; \mathrm{if} \, y_i = -1 \\[0.2em]
		\frac{1}{4} \, T \, (3 + s_i^+ + s_i^- - s_i^+ s_i^-) \; \mathrm{if} \, y_i =  0 \\[0.2em]
		\frac{1}{4} \, T \, (3 - s_i^+ + s_i^- + s_i^+ s_i^-) \; \mathrm{if} \, y_i = +1 \\
	\end{array}
\end{equation}

and generates the combinatorial value table of \cref{tab:c_tc_s_combinatorial_value_table}.

\begin{table}
    \caption{Combinatorial value table for $C_{\text{TC}}(\mathbf{s})$}
    \centering
    \begin{tabular}{|c|c c c c|}
    	\hline
	    Previous & \multicolumn{4}{c|}{$C_{\text{TC}}(\mathbf{s})$ given $(s_i^-, s_i^+)$ and resulting Position $z_i$} \\
        Position & $(-1, -1)$ & $(-1, +1)$ & $(+1, -1)$ & $(+1, +1)$ \\
        $y_i$ & $z_i=0$ & $z_i=+1$ & $z_i=-1$ & $z_i=0$ $\dagger$ \\
	    \hline
        $-1$ & $T$ & $T$ & $0$ & $T$ \\
        $ 0$ & $0$ & $T$ & $T$ & $T$ \\
        $+1$ & $T$ & $0$ & $T$ & $T$ \\
    	\hline
    	\multicolumn{5}{c}{} \\[-0.5em] 
    	\multicolumn{5}{c}{$\dagger$ Degenerate solution of trading long and short is penalized.} \\
    \end{tabular}
    \label{tab:c_tc_s_combinatorial_value_table}
\end{table}

This conditional expression can be converted into an unconditional expression by designing coefficient multipliers based on $y_i$ that provide equivalent behavior

\begin{equation}
	\label{eq:c_tc_s}
	\begin{alignedat}{2}
		& C_{\text{TC}}(\mathbf{s}) = \, & \tfrac{1}{4} \, T \, \big( 3 + (1 - y_i^2 - y_i) s_i^+ + (1 - y_i^2 + y_i) s_i^- \\
		&                                &                                              + \, (2 y_i^2 - 1) s_i^+ s_i^- \big)
	\end{alignedat}
\end{equation}

\subsection{Investment constraint encoding (soft constraint form)}
\label{sec:portfolio_rebalancing_formulation_investment_constraint_encoding_soft}

The investment constraint of \cref{eq:investment_constraint} can be converted to a penalty function using \cref{eq:penalty_functions_for_soft_constraints}, as

\begin{equation}
	\label{eq:p_inv_z}
    P_{\text{INV}}(\mathbf{z}) = A \left( \sum_{i=1}^N z_i - D \right)^2
\end{equation}

where the penalty scaling coefficient $A$ is determined experimentally.

Substitution of the spin variable formulation \cref{eq:risk_return_substitution} into \cref{eq:p_inv_z} yields the spin-system investment penalty function

\begin{equation}
	\label{eq:p_inv_s}
	\begin{alignedat}{2}
		& P_{\text{INV}}(\mathbf{s}) = \, & \frac{A}{4} \sum_{i=1}^N \sum_{j=1}^N (s_i^+ s_j^+ - s_i^+ s_j^- - s_i^- s_j^+ + s_i^- s_j^-) \\
		&                                 & - A D \sum_{i=1}^N (s_i^+ - s_i^-) + A D^2
	\end{alignedat}
\end{equation}

\subsection{Investment constraint encoding (hard constraint form)}
\label{sec:portfolio_rebalancing_formulation_investment_constraint_encoding_hard}

The investment constraint of \cref{eq:investment_constraint} can be realized using a pair of parity ring mixers \cite{Hadfield2019} acting independently on the long and short decision variables of \cref{eq:long_short_decision_variables}. Entanglement between the long and short positions is used to further constrain the set of feasible outcomes to those that have the correct net parity, and thus the correct net investment.

This configuration is designed beginning with the observation that, given a system of $N$ assets encoded as \cref{eq:risk_return_substitution} and a net investment of $D$ as per \cref{eq:investment_constraint}, there are $K$ discrete parity bands where the long and short positions can correctly net off, calculated as

\begin{equation}
	\label{eq:parity_band_count}
	K = N - D + 1
\end{equation}

The design objective becomes to ensure that the long and short mixers can exchange in the subspace of any of these parity bands, but only in a way where outcomes are consistent with the net investment constraint. This is achieved using a combination of computational basis states that establish the minimum long position of $D$ lots, and Bell states whose entanglement assures the required parity relationship across the remaining assets. The Bell states ensure that each additional long position in the solution is accompanied by a short position somewhere in the portfolio.

As in \cite{Hadfield2019}, only one feasible initial state $|\psi_0\rangle$ needs to be prepared this way, as

\begin{equation}
	\label{eq:hard_constraint_psi}
	|\psi_0\rangle = \big( |01\rangle \big)^{\otimes D} \otimes \big( \tfrac{1}{\sqrt{2}} |00\rangle + \tfrac{1}{\sqrt{2}} |11\rangle \big)^{\otimes(N-D)}
\end{equation}

where qubit state pairs are represented as $|x_i^- x_i^+\rangle$.

This configuration provides a feasible initial state whose preparation is likely to be efficient on NISQ hardware. However, it does not provide a uniform solution probability across parity bands. Instead, a binomial distribution is created based on the population of possible solutions, with probability of measurement

\begin{equation}
	\label{eq:hard_constraint_probability}
	\mathrm{Pr}(|\{x_i^- = 1\}| = k) = \frac{\binom{N-D}{k}}{2^{(N-D)}}
\end{equation}

where \begin{itemize}
	\item $k \in \mathbb{Z} \mid 0 \leq k \leq K$ is the parity band
	\item $|\{x_i^- = 1\}|$ is the count of shorted assets
\end{itemize}

We acknowledge the potential for solution bias due to this initial configuration. We consider it an exercise for future development on NISQ hardware to consider this in concert with issues of circuit depth, as any improved solution will need to come within the coherence time of the target hardware.

\begin{figure}
	\centering
	\includegraphics[width=\linewidth]{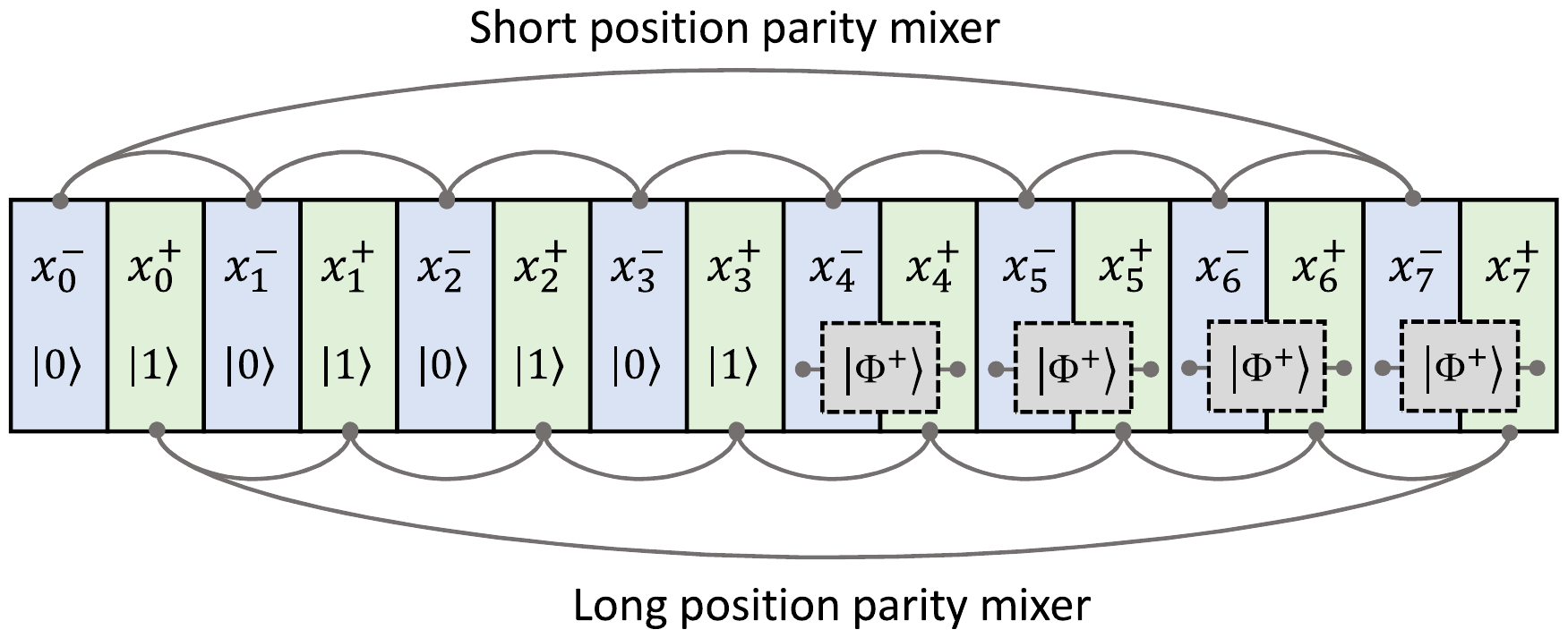}
	\caption{Initial state for the hard constraint form of the investment constraint. The two parity mixers are depicted together with the Bell state entanglement for an example where $N=8$, $D=4$, $K=5$.}
	\label{fig:example_portfolio_rebalancing_mixers}
\end{figure}

A depiction of the initial state and algorithm configuration is provided in \cref{fig:example_portfolio_rebalancing_mixers}. For the same problem size, the initial parity band probability distribution is provided in \cref{tab:example_hard_constraint_parity_bands}.

\begin{table}
    \caption{Parity bands for the hard constraint form of the investment constraint. Position counts and initial state probabilities are calculated for an example where $N=8$, $D=4$, $K=5$.}
    \centering
    \begin{tabular}{|c|c c|c|}
    	\hline
	    Parity & \multicolumn{2}{c|}{Position Count} & Initial \\
        Band & (long) & (short) & Probability \\
        $k$ & $|\{x_i^+ = 1\}|$ & $|\{x_i^- = 1\}|$ & $\mathrm{Pr}(|\{x_i^- = 1\}| = k)$ \\
	    \hline
        $0$ & $4$ & $0$ & $6.25\%$ \\
        $1$ & $5$ & $1$ & $25.0\%$ \\
        $2$ & $6$ & $2$ & $37.5\%$ \\
        $3$ & $7$ & $3$ & $25.0\%$ \\
        $4$ & $8$ & $4$ & $6.25\%$ \\
    	\hline
    \end{tabular}
    \label{tab:example_hard_constraint_parity_bands}
\end{table}

%

\subsection{Soft constraint formulation using the Quantum Approximate Optimization Algorithm}
\label{sec:portfolio_rebalancing_formulation_qaoa_realization_soft_constraints}

The portfolio rebalancing application can be realized using an unconstrained optimization approach by combining the spin-system risk-return cost function (\cref{eq:c_rr_s}), trading cost function (\cref{eq:c_tc_s}) and investment penalty function (\cref{eq:p_inv_s}), as

\begin{equation}
	\label{eq:c_soft_s}
	C_{\text{soft}}(\mathbf{s}) = C_{\text{RR}}(\mathbf{s}) + C_{\text{TC}}(\mathbf{s}) + P_{\text{INV}}(\mathbf{s})
\end{equation}

Solving this formulation using the Quantum Approximate Optimization Algorithm involves its execution as described in \cref{sec:preliminaries_quantum_approximate_optimization}. The QAOA cost operator is of the \cref{eq:qaoa_ising_cost_operator} form, generated by substitution of the Pauli-Z operator $\sigma_i^z$ for each spin $s_i$ in \cref{eq:c_soft_s}. The QAOA unconstrained mixing operator is of the standard \cref{eq:qaoa_farhi_mixing_operator} form.

\subsection{Hard constraint formulation using the Quantum Alternating Operator Ansatz}
\label{sec:portfolio_rebalancing_formulation_qaoa_realization_hard_constraints}

The portfolio rebalancing application can also be realized using a constrained optimization approach by combining the spin-system risk-return cost function (\cref{eq:c_rr_s}) and trading cost function (\cref{eq:c_tc_s}) as

\begin{equation}
	\label{eq:c_hard_s}
	C_{\text{hard}}(\mathbf{s}) = C_{\text{RR}}(\mathbf{s}) + C_{\text{TC}}(\mathbf{s})
\end{equation}

and whose investment constraint is realized as described in \cref{sec:portfolio_rebalancing_formulation_investment_constraint_encoding_hard} with an entangled pair of parity mixers.

Solving this formulation using the Quantum Alternating Operator Ansatz involves its execution as described in \cref{sec:preliminaries_quantum_approximate_optimization}. The QAOA cost operator remains of the \cref{eq:qaoa_ising_cost_operator} form, generated by substitution of the Pauli-Z operator $\sigma_i^z$ for each spin $s_i$ in \cref{eq:c_hard_s}. The two QAOA constrained mixing operators are of the \cref{eq:qaoa_parity_mixing_operator} form, with one mixer applying to the set of long position spin variables $\lbrace s_i^+ \rbrace$, the other applying to the set of short position spin variables $\lbrace s_i^- \rbrace$, and the initial feasible state being as \cref{eq:hard_constraint_psi}.


\section{Portfolio Rebalancing Experimental Results}
\label{sec:portfolio_rebalancing_experimentation}

We execute the portfolio rebalancing formulation upon an idealized simulator of a gate-model quantum computer to assess its tractability prior to execution on NISQ hardware. We do this in a sequence of steps designed to verify the application is implemented correctly.

\subsection{Experimental data}

The input data for this experiment is the daily returns for share prices on the Australian ASX.20 market in 2017. The data covered 20 stocks and 252 trading days. Summary statistics for the daily returns are presented in \cref{tab:daily_average_asset_returns}.

\begin{table}
	\caption{Daily average returns for ASX.20 in 2017.}
	\centering
	\begin{tabular}{ |c c c|c c c| }
		\hline
		Stock & Average & S.D. & Stock & Average & S.D. \\
		& ($\mu_i$) & ($\sigma_i$) & & ($\mu_i$) & ($\sigma_i$) \\
		\hline
        AMP &  0.000401 & 0.009988 & QBE & -0.000316 &  0.014433 \\
        ANZ &  0.000061 & 0.010024 & RIO &  0.001230 &  0.014854 \\
        BHP &  0.000916 & 0.013465 & SCG & -0.000176 &  0.010974 \\
        BXB & -0.000619 & 0.015910 & SUN &  0.000396 &  0.010007 \\
        CBA &  0.000212 & 0.009201 & TLS & -0.000881 &  0.013377 \\
        CSL &  0.001477 & 0.013156 & WBC &  0.000184 &  0.009907 \\
        IAG &  0.001047 & 0.011216 & WES &  0.000492 &  0.008399 \\
        MQG &  0.000794 & 0.010052 & WFD &  0.000291 &  0.013247 \\
        NAB &  0.000204 & 0.009193 & WOW &  0.000674 &  0.008477 \\
        ORG &  0.001500 & 0.014958 & WPL &  0.000491 &  0.010873 \\
		\hline
	\end{tabular}
	\label{tab:daily_average_asset_returns}
\end{table}

\begin{figure}
	\centering
	\includegraphics[width=\linewidth]{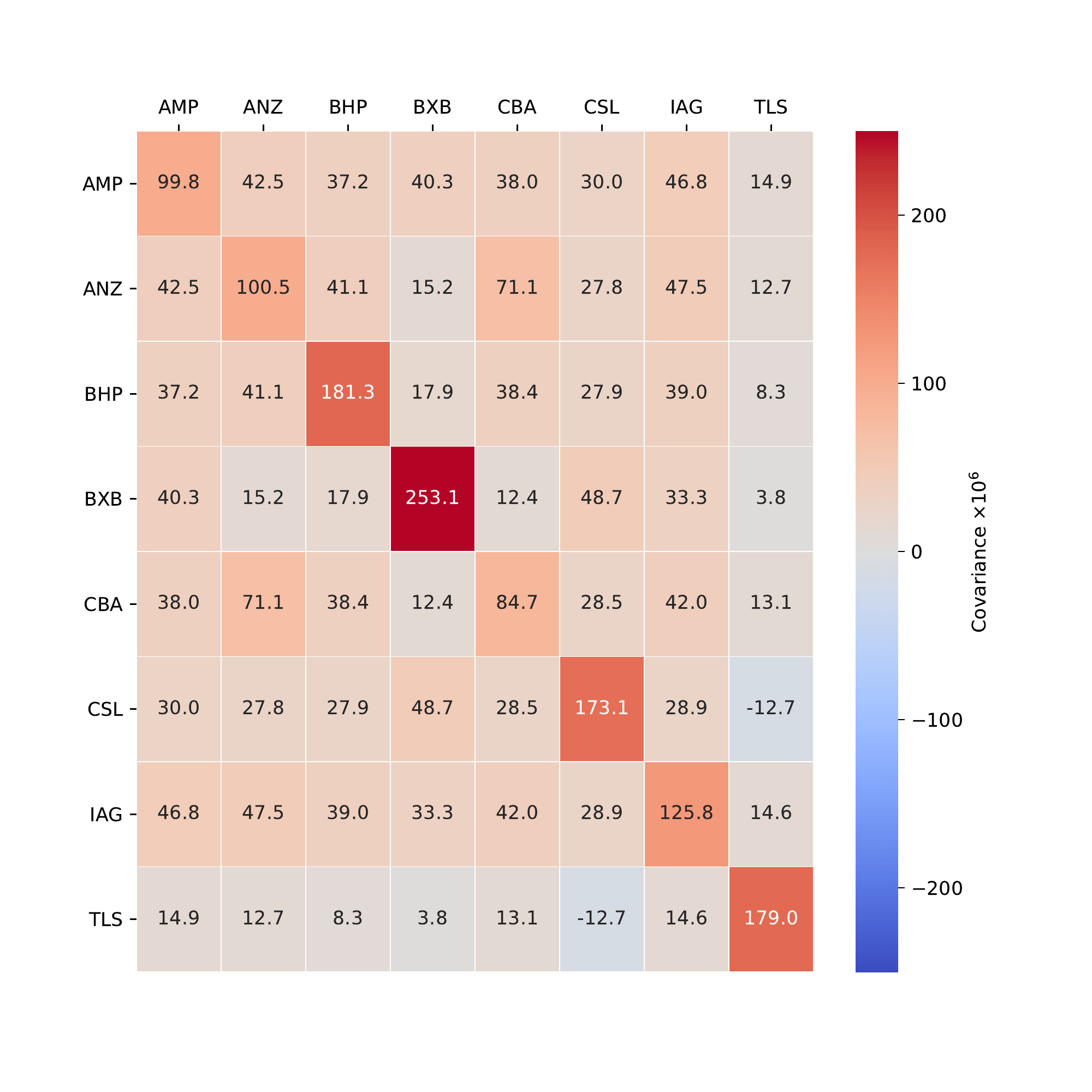}
	\caption{Daily average asset covariance as $\sigma_{ij} \times 10^6$ for 8 stocks selected from the ASX.20 in 2017}
	\label{fig:daily_average_asset_covariance}
\end{figure}

Data for $N = 8$ stocks (AMP, ANZ, BHP, BXB, CBA, CSL, IAG, MQG) were selected for use in the experiments. Daily average returns ($\mu_i$) can be found in \cref{tab:daily_average_asset_returns}. The asset returns covariance ($\sigma_{ij}$) can be found in \cref{fig:daily_average_asset_covariance}. By convention, covariance $\sigma_{ii} = \sigma_i^2$ where $\sigma_i$ is the standard deviation (S.D.).

For all experiments, a target holding of $D = 4$ discrete lots is used, as per \cref{eq:investment_constraint}.

\subsection{Calculation of penalty scaling}

A simple method to calculate penalty scaling coefficient $A$ in \cref{eq:p_inv_z} was adopted, as

\begin{equation}
	\label{eq:A_penalty_max_min}
    A > \text{max}\left[ C(\mathbf{s}) \right] - \text{min}\left[ C(\mathbf{s}) \right]
\end{equation}

where $C(\mathbf{s}) = C_{\text{RR}}(\mathbf{s}) + C_{\text{TC}}(\mathbf{s})$ is the unconstrained part of the soft constraint formulation of \cref{eq:c_soft_s}. This ensures that the value of $C_{\text{soft}}(\mathbf{s})$ for any unfeasible solution is greater than the energy for all feasible solutions. The performance of this setting for realizing feasible solutions will be validated experimentally.

\subsection{Investigation of the solution space}

We begin by evaluating the solution space using brute force methods. Given $N = 8$ stocks, the cost function encoding requires 16 spin variables, and presents a combinatorial space of $2^{16} = 65536$ solution states. Of these, only 1820 (2.78\%) are feasible, determined by the investment constraint of \cref{eq:investment_constraint} for $D = 4$.

\begin{figure}
	\centering
	\includegraphics[width=\linewidth]{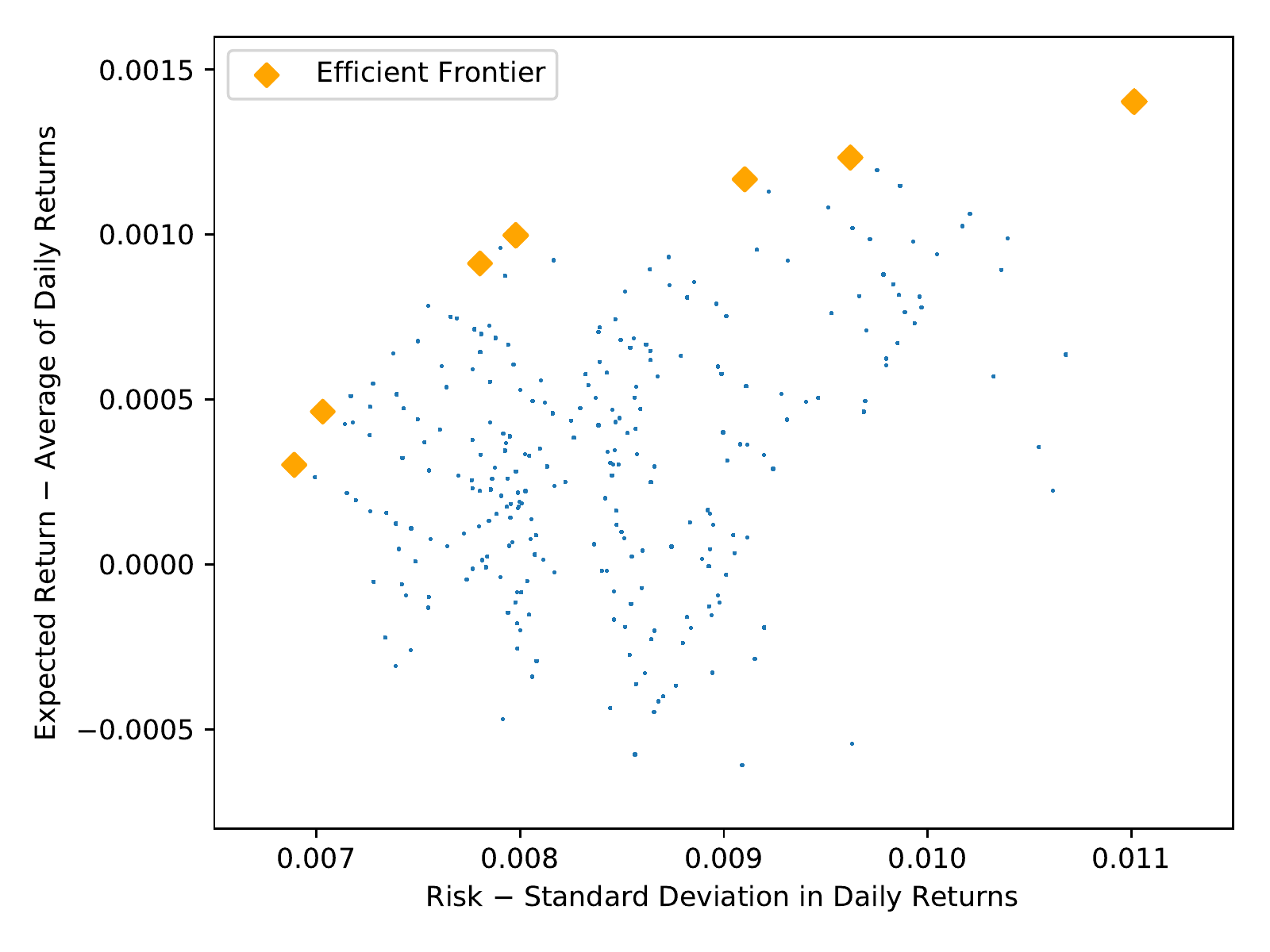}
	\caption{All feasible solutions to $C_{\text{RR}}(\mathbf{z})$ plotted as expected returns versus risk. Optimal solutions for varying $\lambda$ define the discrete efficient frontier.}
	\label{fig:efficient_frontier}
\end{figure}

In finance, it is common to plot feasible portfolio solutions against the accumulated values of expected return and risk from \cref{eq:risk_return_function}\footnote{For clarity, expected return is $\sum \mu_i z_i$, while risk is $\sqrt{\sum \sigma_{ij} z_i z_j}$.}, independent of risk-return control parameter $\lambda$. This is illustrated in \cref{fig:efficient_frontier}. As $\lambda$ is varied, the optimal discrete solution point changes, creating a discrete efficient frontier of available portfolio solutions.

\subsection{Investigation of a single QAOA circuit}

\begin{figure}
	\centering
	\includegraphics[width=\linewidth]{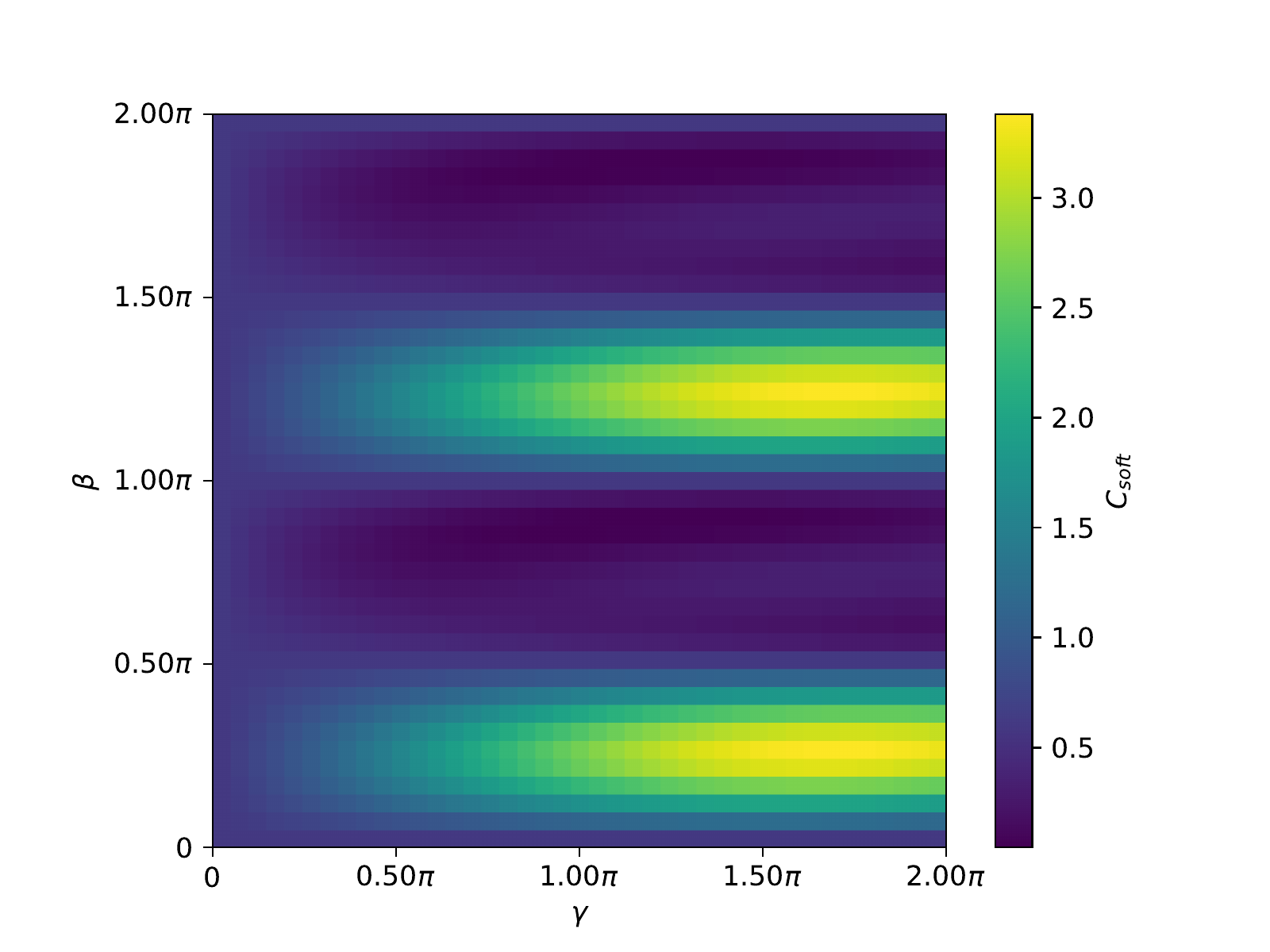}
	\caption{$\langle \psi_1 | C_{\text{soft}}(\mathbf{s}) | \psi_1 \rangle$ using input data based on average daily returns, with $\lambda = 0.9$, $A = 0.03$, $T = 0$, $p = 1$.}
	\label{fig:beta_gamma_experiment_1}
\end{figure}

Before enabling the outer classical optimization loop of QAOA, we first investigated the behavior for a single iteration of the Quantum Approximate Optimization Algorithm circuit on formulation \cref{eq:c_soft_s}. The penalty scaling coefficient is calculated using \cref{eq:A_penalty_max_min} as $A = 0.03$. The risk-return control parameter is set as $\lambda = 0.9$, and trading cost $T = 0$. We vary $\beta$ and $\gamma$ angles for a $p = 1$ execution of \cref{eq:qaoa_final_state}, and obtain the results in \cref{fig:beta_gamma_experiment_1}.

\begin{figure}
	\centering
	\includegraphics[scale=0.4286]{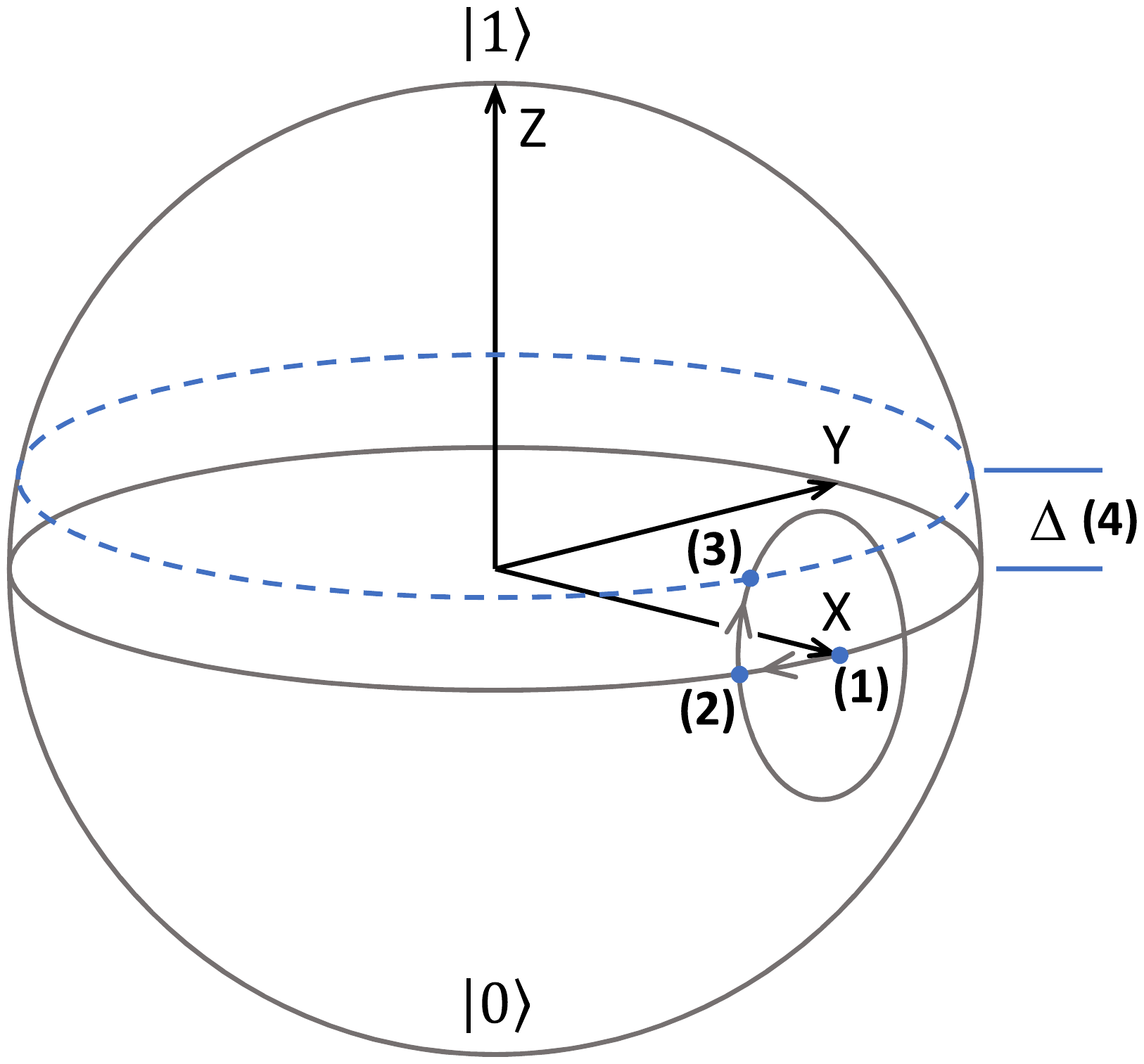} 
	\caption{Effect of poor scaling in QAOA cost function $C$: (1) State is initialized to $| \psi_0 \rangle = | + \rangle ^ {\otimes N}$; (2) $U(C, \gamma) = e^{-i \gamma C}$ performs a small $z$-rotation when $C \ll 1$; (3) $U(B, \beta) = e^{-i \beta B}$ performs an $x$-rotation to produce $| \psi_1 \rangle$; which (4) results in a small $\Delta$ change in expectation $\langle \psi_1 | C_{\text{soft}}(\mathbf{s}) | \psi_1 \rangle$.}
	\label{fig:effect_of_poor_scaling}
\end{figure}

Slow variation in expectation value is observed for changes in $\gamma$. This occurs due to a problem with cost function scaling. For effective mixing of the initial superposed states $| \psi_0 \rangle$ in \cref{eq:qaoa_initial_state}, the cost function $C$ must be scaled close to unity. If $C$ is small, then $\gamma C$ is small, and the $z$-rotation of $U(C, \gamma) = e^{-i \gamma C}$ is also small\footnote{We acknowledge that for multiple rotations upon the same spin, the angles may accumulate. However in this experiment, the unitary count is $\mathcal{O}(N)$ with $N = 8$ and does not significantly alter the scale.}. This has the effect of a small change on expectation value for $p = 1$ as illustrated in \cref{fig:effect_of_poor_scaling}, and slow mixing to the complete solution space for larger $p$.

\begin{figure}
	\centering
	\includegraphics[width=\linewidth]{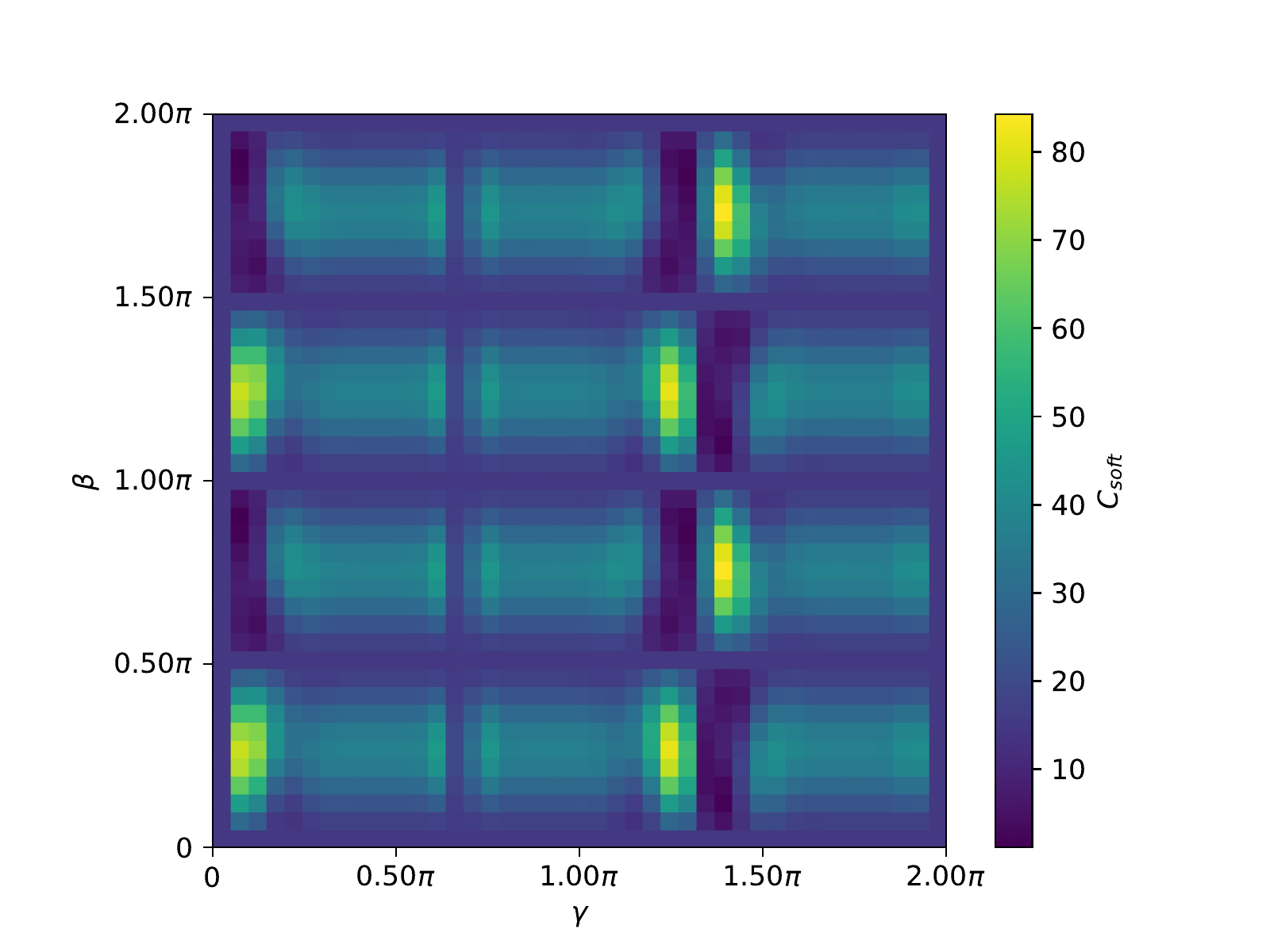}
	\caption{$\langle \psi_1 | C_{\text{soft}}(\mathbf{s}) | \psi_1 \rangle$ using input data based on annualized returns, with $\lambda = 0.9$, $A = 0.75$, $T = 0$, $p = 1$.}
	\label{fig:beta_gamma_experiment_2}
\end{figure}

We address this issue by scaling the input data to annualized returns instead of daily returns. Annualized returns are calculated from daily returns by multiplying both the average $\mu_i$ and covariance $\sigma_{ij}$ by 250. The penalty scaling coefficient is recalculated as $A = 0.75$. Running the experiment again we obtain the results in \cref{fig:beta_gamma_experiment_2}. Initial mixing is observed to be improved, evidenced by the periodic structure in the data.

\subsection{Evaluation of QAOA algorithms for a single portfolio}

We enable the outer classical optimization loop for both the Quantum Approximate Optimization Algorithm and Quantum Alternating Operator Ansatz formulation of the problem. We investigate the cost function statistics for solutions found by these algorithms as an initial representation of the value of the resulting portfolio, for a range of quantum iterations $p$. We include brute force baseline statistics, generated from a uniform distribution across all possible and all feasible solutions. All statistics for QAOA experiments are calculated across 20 randomly seeded simulated runs.

\begin{figure}
	\centering
	\includegraphics[width=\linewidth]{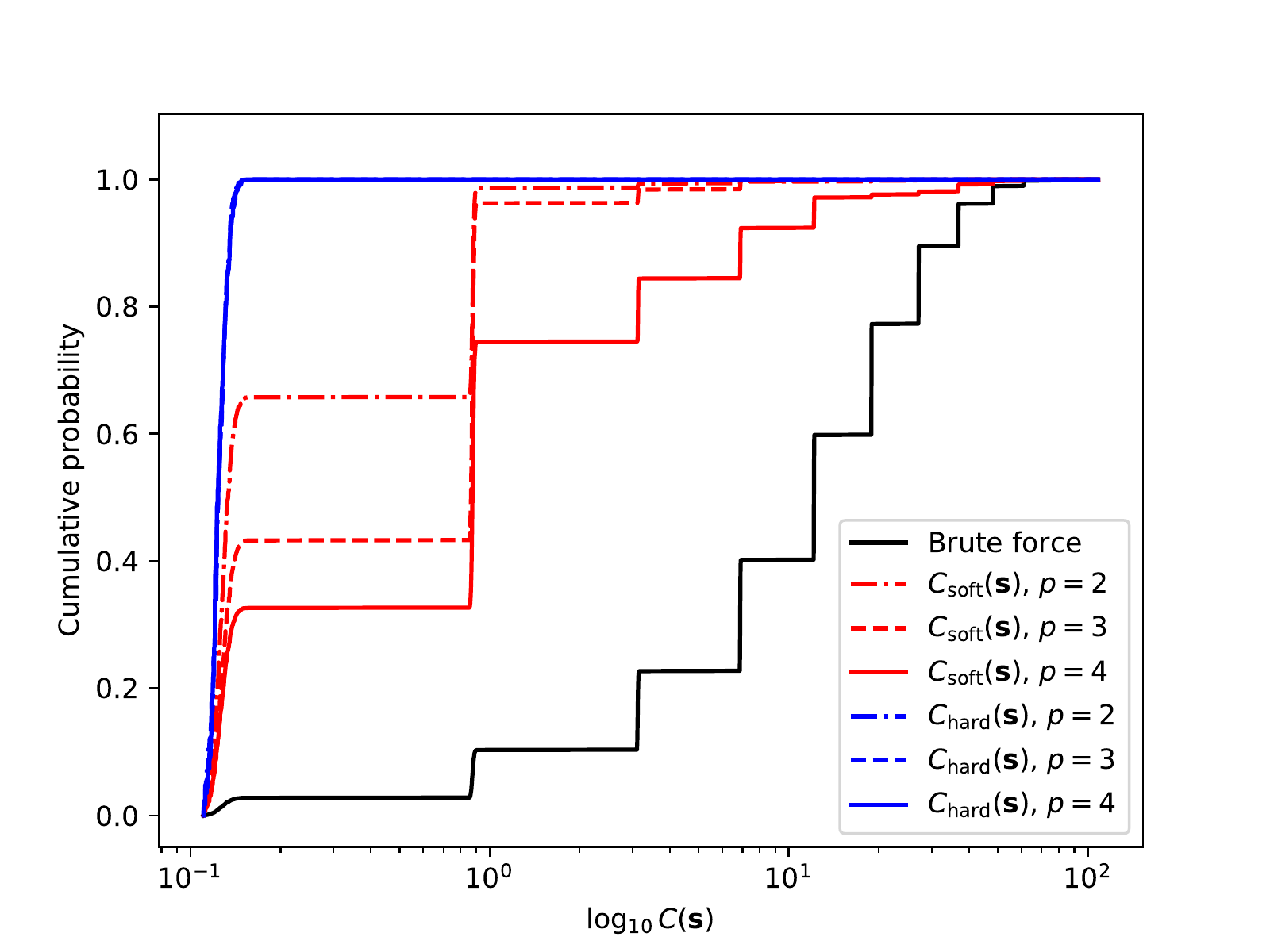}
	\caption{Cumulative probability of $C(\mathbf{s})$ for \textbf{all possible} (i) \enquote{Brute force} solutions to $C(\mathbf{s})$; (ii) Quantum Approximate Optimization Algorithm solutions to $C_{\text{soft}}(\mathbf{s})$; and (iii) Quantum Alternating Operator Ansatz solutions to $C_{\text{hard}}(\mathbf{s})$; with $\lambda = 0.9$, $A = 0.75$, $T = 0$.}
	\label{fig:cumulative_probability_1}
	
	\includegraphics[width=\linewidth]{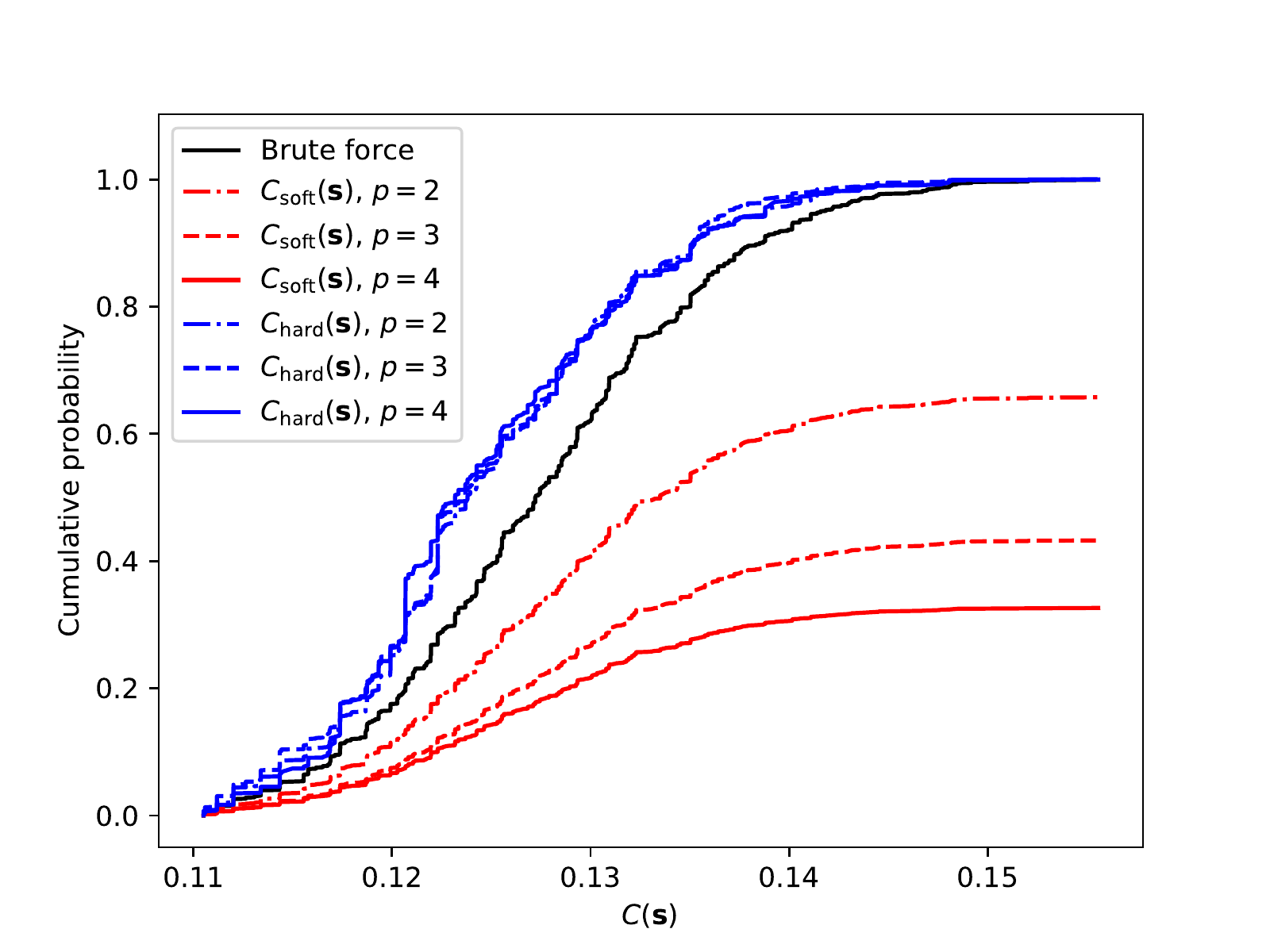}
	\caption{Cumulative probability of $C(\mathbf{s})$ for \textbf{all feasible} (i) \enquote{Brute force} solutions to $C(\mathbf{s})$; (ii) Quantum Approximate Optimization Algorithm solutions to $C_{\text{soft}}(\mathbf{s})$; and (iii) Quantum Alternating Operator Ansatz solutions to $C_{\text{hard}}(\mathbf{s})$; with $\lambda = 0.9$, $A = 0.75$, $T = 0$.}
	\label{fig:cumulative_probability_2}
\end{figure}

The cumulative probability of the cost function is illustrated in \cref{fig:cumulative_probability_1}. All algorithms show a significant improvement in results compared to random draw from the solution space. The Quantum Approximate Optimization Algorithm returns feasible solutions with between 33\% and 66\% probability, validating the choice of penalty scaling coefficient $A$. Interestingly, performance is best for lower $p$, and may be an artifact of the Nelder-mead optimizer. The Quantum Alternating Operator Ansatz formulation returns feasible solutions with 100\% probability, verifying the correct implementation of the mixing function and the expected simulated behavior.

The cumulative probability of the cost function restricted to the domain of feasible solutions is illustrated in \cref{fig:cumulative_probability_2}. The Quantum Alternating Operator Ansatz shows superior performance with respect to a random selection of feasible solutions, validating that not only is the algorithm assisting with identification of a feasible solution in the large combinatorial space, but that it is optimizing for the cost function within that subspace. The difference in performance to the original Quantum Approximate Optimization Algorithm is significant, highlighting the importance of hard constraint design in solving constrained problems using QAOA.

\subsection{Evaluation of QAOA algorithms for portfolio rebalancing}

We broaden the experiment to use the QAOA algorithms for a multi-period rebalancing scenario. In this scenario we consider average annualized returns calculated independently for each of the first 6 months in 2017, and rebalance the portfolio once per month. Trading costs are included at $T = 0.015$, which is of the same order of magnitude as the annualized returns of the available stocks. The monthly statistics for the annualized returns lead us to set $A = 2.5$ as per \cref{eq:A_penalty_max_min}, due to the greater extremes in minimum and maximum values that occur over the shorter period.

We calculate metrics of interest to traders and quantitative analysts, including the total number of trades, the portfolio returns adjusted for trading costs, and the average risk over the period. We test using both the Quantum Approximate Optimization Algorithm and Quantum Alternating Operator Ansatz formulations, and compare to the optimal trajectory determined by the brute force methods. Note that optimization is performed independently at each time-step, meaning that the optimal trajectory is not necessarily the one discovered by the brute force methods, as multi-period market foresight is not available in this scenario.

\begin{figure}
	\centering
	\includegraphics[width=\linewidth]{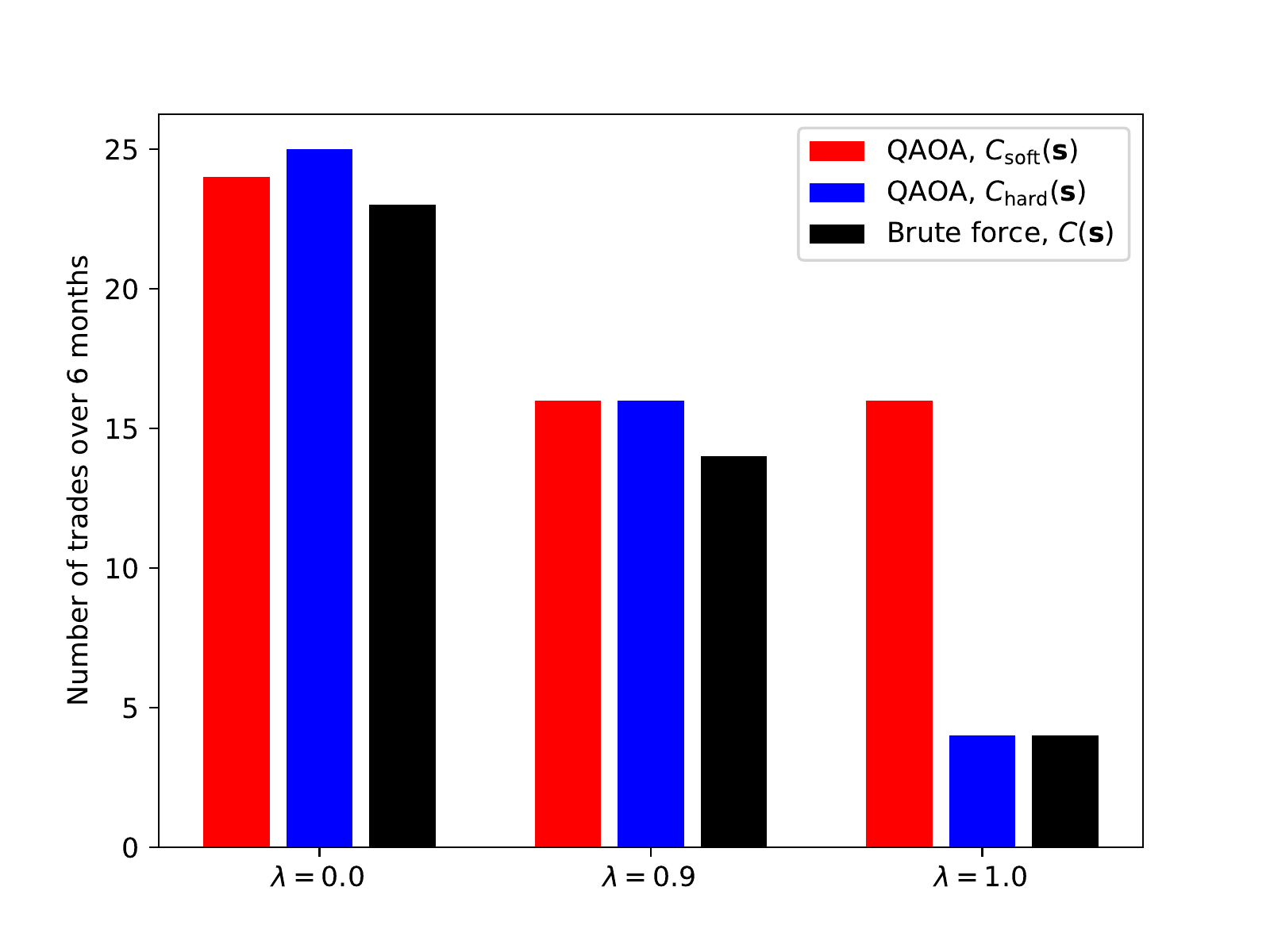}
	\caption{Number of trades over 6 months with $A = 2.5$, $T = 0.015$, $p = 4$.}
	\label{fig:n_trades_versus_lambda}
	
	\includegraphics[width=\linewidth]{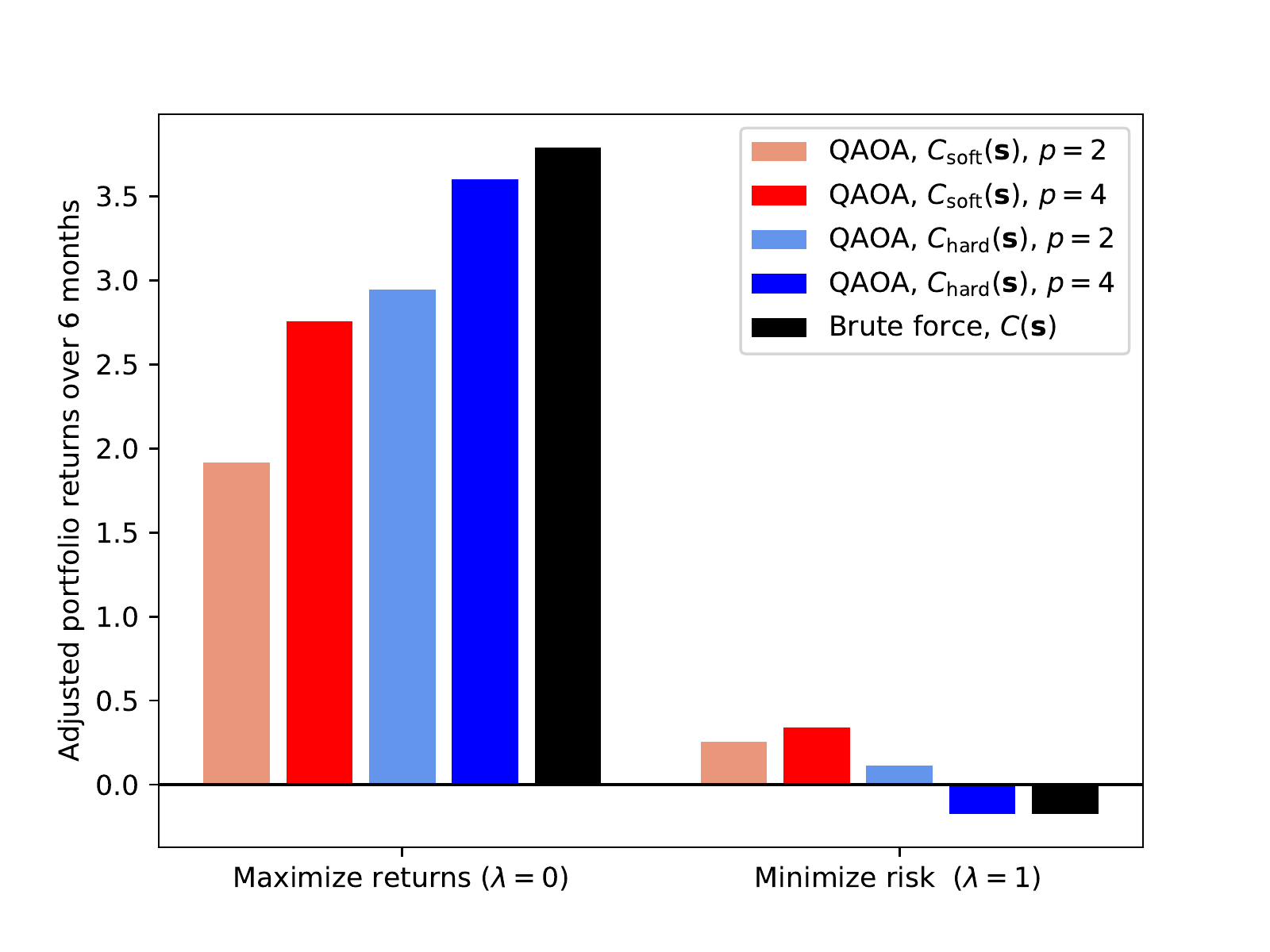}
	\caption{Adjusted returns over 6 months with $A = 2.5$, $T = 0.015$.}
	\label{fig:adj_returns_versus_lambda}

	\includegraphics[width=\linewidth]{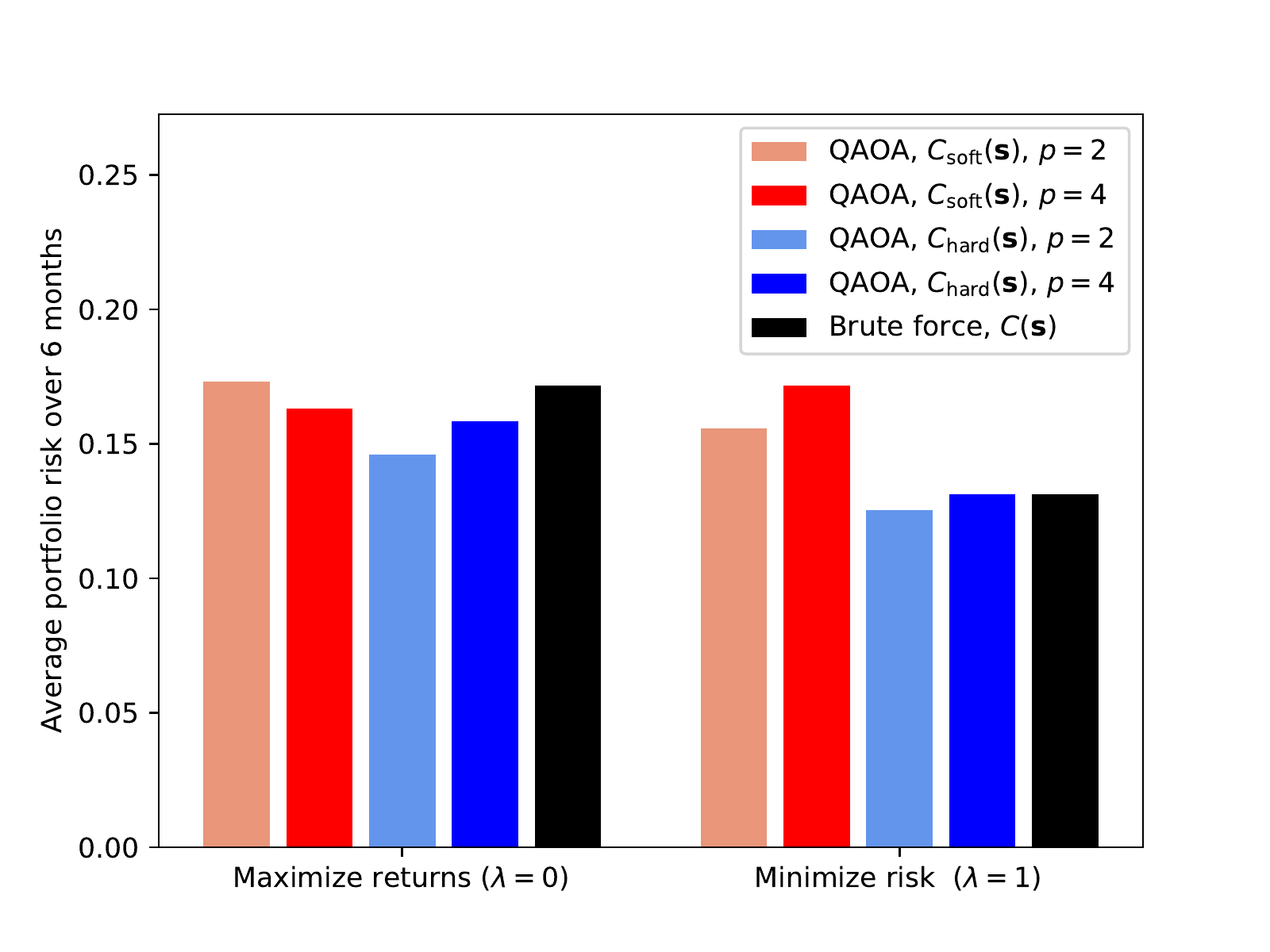}
	\caption{Average risk over 6 months with $A = 2.5$, $T = 0.015$.}
	\label{fig:risk_versus_lambda}
\end{figure}

\cref{fig:n_trades_versus_lambda} shows the total number of trades performed over 6 months of rebalancing. From zero initial holdings, the maximum number of trades to reach the target $D = 4$ is 8 trades; 6 long and 2 short. For the remaining 5 months, the maximum number of trades is 4, netting off changes in long and short positions. This puts an upper limit on the number of trades over 6 months at 28. We observe high volumes of trading for the $\lambda = 0$ case of maximizing returns, consistent with chasing returns month-to-month that is expected when the trading cost is similar to the available returns. As $\lambda$ is increased to favor minimizing risk, trading activity decreases, consistent with the observation in \cref{fig:daily_average_asset_covariance} that there are not very many pairs of stocks with the negative correlation required to reduce risk. Qualitatively both variants of QAOA are observed to perform close to optimal, with the exception of one poor result from the original QAOA formulation.

\cref{fig:adj_returns_versus_lambda} shows the adjusted portfolio returns $\mu \boldsymbol{\cdot} \mathbf{z} - C_{\text{TC}}(\mathbf{z})$ over 6 months of rebalancing. This plot shows the improved performance of the Quantum Alternating Operator Ansatz in achieving good returns when instructed ($\lambda = 0$), measured as within 5\% of the brute force result for this experimental campaign when $p=4$. Superior performance is also observed for $p=4$ over $p=2$ for both QAOA variants, consistent with that expected from deeper circuits with a larger hyper-parameter space. Adjusted returns vary about zero when returns are excluded from the objective.

\cref{fig:risk_versus_lambda} shows the average risk $\sqrt{\mathbf{z}^{\text{T}} \sigma \mathbf{z}}$ over 6 months of rebalancing. This plot again shows an improved performance for the Quantum Alternating Operator Ansatz that, when directed ($\lambda = 1$) to reduce risk, found the optimal solution for each of the 6 months for this experimental campaign when $p=4$. The lower risk for $p=2$ is an artifact of the trading trajectory, where one sub-optimal decision can lead to alternatives with a net benefit.


\section{Conclusion and Future Work}
\label{sec:portfolio_rebalancing_conclusion}

In this work we have shown the application of QAOA to a portfolio rebalancing use case of interest to the financial services industry. We established the characteristics and relevance of the use case and highlighted the potential for quantum computing to impact institutional processes. We formulated two approaches to solving this exemplar application, one using soft constraints and the Quantum Approximate Optimization Algorithm, and another using hard constraints and the Quantum Alternating Operator Ansatz. For the hard constraint formulation, we designed a mixer based on the entanglement of parity mixers across long and short decision variables, which is novel in this space and demonstrates our process for tailoring existing quantum algorithms to industry use cases. We undertook an initial experimental campaign on an idealized simulator of a gate-model quantum computer to verify our implementation and to establish use case tractability.

The experimental results highlighted key issues of effective use of QAOA, including input data scaling. They qualitatively established performance advantages for the Quantum Alternating Operator Ansatz, and performance for both adjusted returns and average portfolio risk, being two key metrics of interest to industry analysts. Experimental analysis demonstrates the potential tractability of this application on Noisy Intermediate-Scale Quantum (NISQ) hardware, identifying portfolios within 5\% of the optimal adjusted returns and with the optimal risk for a small eight-stock portfolio when noise and coherence time limitations are ignored.

These encouraging results warrant further investigation, including (i) circuit depth via hardware resource analysis; (ii) impact of noise via simulated performance analysis; and (iii) validation on current generation NISQ hardware. These steps help uncover the full-stack parameters critical to this use case's trajectory towards potential quantum advantage.

\renewcommand*{\UrlFont}{\rmfamily}  
\printbibliography

\end{document}